\documentclass[a4paper,11pt]{article}
\usepackage{pos}

\title{High-energy Neutrino and Gamma Ray Emission from Clusters-like Perseus}

\author*[a]{Saqib Hussain}
\author[a]{Gabrijela Zaharijas}
\author[b, c]{Klaus Dolag}

\affiliation[a]{University of Nova Gorica, \\ Vipavska 13, SI-5000 Nova Gorica, Slovenia.}
\affiliation[b]{
University Observatory Munich, Scheinerstr 1, D-81679 Munchen, Germany.}
\affiliation[c]{Max Planck Institute for Astrophysics, Karl-Schwarzschild-Str 1, D-85741 Garching, Germany.}

\emailAdd{saqib.hussain@ung.si}
\emailAdd{gabrijela.zaharijas@ung.si}

\abstract{
We calculate the high-energy gamma-ray and neutrino emissions from galaxy clusters like Perseus that host active galactic nuclei (AGNs). Our primary objective is to distinguish the emission from the central source, such as NGC$1275$, from the diffuse emission originating in the outskirts of the Perseus cluster. Due to unique magnetic-field configuration, CRs with energy $\leq 10^{17}$ eV can be confined within these structures over cosmological time scales, and generate secondary particles, including neutrinos and gamma-rays, through interactions with the background gas and photons. We employ three-dimensional cosmological magnetohydrodynamical simulations of structure formation to model the turbulent intracluster medium (ICM). We propagate CRs in intracluster medium (ICM) and intergalactic medium using multi-dimensional Monte Carlo simulations, considering all relevant photohadronic, photonuclear, and hadronuclear interactions. We also include the cosmological evolution of sources like Perseus. By comparing our results with the existing upper limits from IceCube for galaxy clusters and the sensitivity of CTA, we predict that these observatories could potentially establish a new class of astrophysical sources capable of emitting high-energy multi-messenger signals. We also compute the contribution from clusters like Perseus to the diffuse neutrino and gamma-ray background.
}

\ConferenceLogo{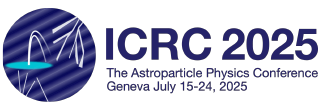}

\FullConference{39th International Cosmic Ray Conference (ICRC2025)\\
 15–24 July 2025\\
Geneva, Switzerland\\}

\begin{document}
\maketitle

\section{Introduction}

The origin of diffuse gamma-rays, neutrinos, and CRs is unknown. The most probable astrophysical sources that can produce these messengers are AGNs \citep[][see the references therein]{ajello2015origin, di2013diffuse} and galaxy clusters hosting AGNs \citep{fang2018linking, hussain2021high, hussain2023diffuse}.
Using Fermi-LAT data, previous studies (see, e.g., \citep{Raniere2022isotropic}) have shown that active galactic nuclei (AGNs) can account for at most about $40\%$ of the diffuse gamma-ray background. Associating the diffuse neutrino flux with AGN is even more uncertain and remain an open problem.
The fact that the magnitude of fluxes of high-energy cosmic messengers is comparable poses a proposition of common origin, i.e., a single class of sources can produce them.
In light of the preceding arguments, galaxy clusters are the most plausible candidates for producing cosmic messengers, including CRs, gamma-rays, and neutrinos.
Given the extreme environment, large size ($\sim 2$ Mpc), and high magnetic field strength ($\sim 1\mu$ G), clusters can confine high-energy ($\leq 10^{18}$~eV) CRs up to the age of the Universe. 
During their confinement, CRs can interact with the ICM gas and photon fields to produce neutrinos and gamma-rays.
%

In this work, we study the production of high-energy gamma-rays and neutrinos from galaxy clusters. Our primary focus is to explore the Perseus cluster because it is the brightest X-ray source \citep{UrbanXray2014, TNGCluster2024} in the local Universe and probably can produce these messengers. The upcoming CTA observatory will focus specifically on the Perseus cluster, establishing whether it can emit gamma-rays or not. This aspect underscores the relevance and timeliness of our work.

To study the diffuse emissions from the galaxy clusters, we used the most detailed numerical method, combining three-dimensional magnetohydrodynamic (MHD) simulations with multi-dimensional Monte-Carlo simulations \citep[][]{hussain2021high, hussain2023diffuse}.
The MHD simulation probes the background ICM, including gas and magnetic field distributions, while Monte-Carlo simulations are used to study the CR propagation in the ICM and intergalactic medium.
The results are presented in the next section, and the last section is dedicated to discussion and conclusions.

\section{Results}

To probe the ICM, we explore the MHD simulation \citep{SLOWDolag2023} of the local Universe \footnote{https://www.usm.uni-muenchen.de/~dolag/Simulations/} which is designed to investigate anomalies in the local Universe up to $500\, h^{-1}\, \text{Mpc}$. 
The initial conditions for the simulation are derived from peculiar velocities based on the CosmicFlow catalog \citep{cosmicoutflow4}.
Key structures in this local web simulation are massive galaxy clusters such as Virgo, Coma, Perseus, Norma, Centaurus, and Hercules.
We focused on the Perseus cluster, and the gas and magnetic field profiles are directly obtained from the MHD simulation.

We study the propagation of CRs through the Perseus cluster using multi-dimensional Monte-Carlo simulations.
The CRPropa code \citep{batista2016crpropa} \footnote{https://crpropa.github.io/CRPropa3/} is used to investigate the propagation of CRs and gamma-rays inside the cluster, as well as through the intergalactic medium. This code allows for the use of user-defined magnetic field and density profiles, which in this case are obtained from MHD simulations.
We considered all the relevant CR and gamma-ray interactions, namely: proton-
proton (pp) interactions, photopion production, Bethe-Heitler pair
production, pair production (single, double, triplet), and inverse Compton scattering (ICS). 
The background photon fields considered in this work are: CMB, EBL, and radio.

In this work, we assume that CRs consist of only protons because photodisintegration dominates the pion production in the case of heavy ions. Therefore, our assumption is quite reasonable \cite[see e.g.,][]{kotera2009propagation}.  The gamma-rays produced in this scenario are purely hadronic. 
During the propagation of gamma-rays in the intergalactic medium,
the magnetic field is not considered, as it is highly uncertain, and most likely, it would not make any significant changes in the results, especially for energy above $10$ GeV.

\begin{figure}[ht!]
\centering
\includegraphics[width=0.6\textwidth]{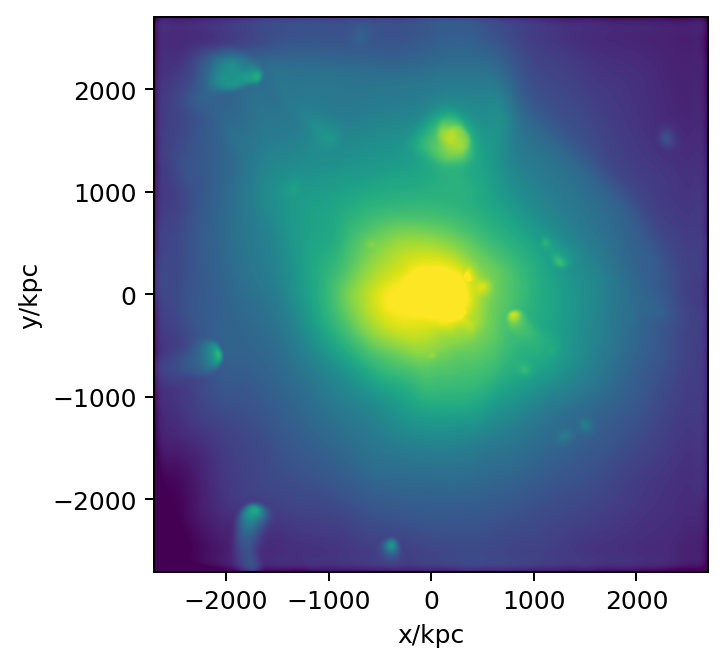}
\caption{2D map of gas distribution of the Perseus cluster \citep{SLOWDolag2023}.   
}\label{fig:Perseus_gas}
\end{figure}

In Fig. \ref{fig:Perseus_gas}, we present the two-dimensional (2D) map of the gas distribution of the Perseus cluster, obtained from the MHD simulation \citep{SLOWDolag2023}. 

\begin{figure}[ht!]
\centering
\includegraphics[width=0.7\textwidth]{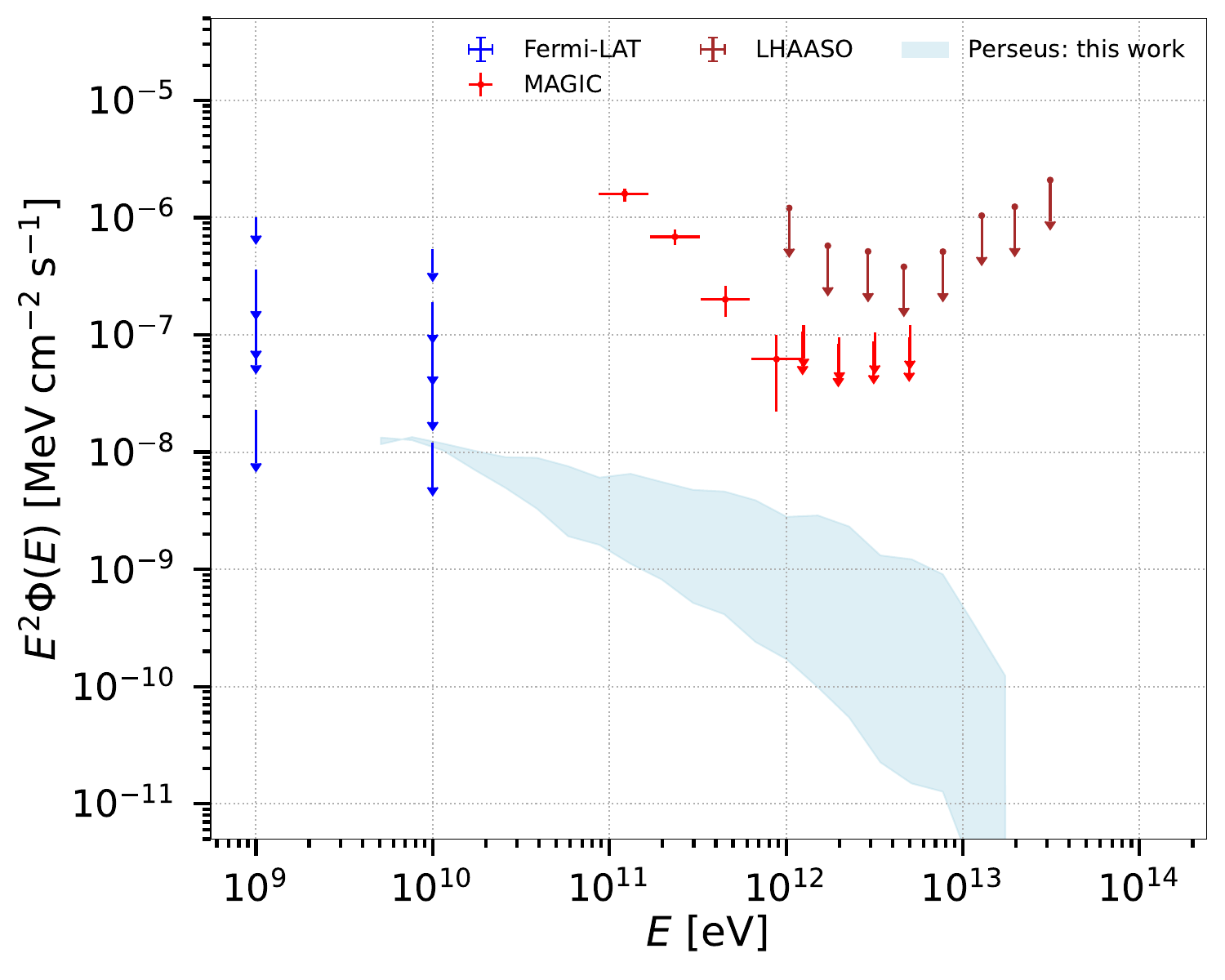}
\caption{Gamma-ray picture of Perseus cluster.
Blue band represents our results for parameters $\alpha= 2.0 - 2.5\, \text{and} \, E_\text{max}=10^{16}- 10^{17} \text{eV}$. The observations and upper limits for the Perseus cluster by experiments such as MAGIC (red error bars) \citep{MAGIC2016NGC1275} and LHAASO (brown arrows) \citep{di2016lhaaso} are depicted, respectively. The Fermi-LAT upper limits for massive clusters (blue arrows, top to bottom: A400, A3112, A1367, Coma, EXO0422) \citep{FermiCluster2014search} are also shown.
\label{fig:Perseus_gamma}
}
\end{figure}

Fig. \ref{fig:Perseus_gamma}, shows the flux of gamma-rays for the Perseus cluster. We plotted the results for the parameters $\alpha= 2.0 - 2.5\, \text{and} \, E_\text{max}=10^{16}- 10^{17} \text{eV}$. For normalization, we considered that $1\, \%$ of the thermal energy of the cluster goes to CRs.
The ratio $X_{CR} = E_\text{CR}/E_\text{thermal} \approx 0.01$, where $E_\text{CR}$ is the integral CR energy above $1$~GeV, and $E_\text{thermal}$ is the total thermal energy of the cluster.
We also plotted upper limits of gamma-rays for the Perseus cluster provided by experiments such as MAGIC \citep{MAGIC2016NGC1275} and Large High Altitude Air Shower Observatory (LHAASO) \citep{LHAASO2025ClusterUL}.
The gamma-ray flux we obtained for the Perseus cluster is well below the upper-limits predicted by MAGIC \citep{MAGIC2016NGC1275} and LHAASO \citep{LHAASO2025ClusterUL}. 
The MAGIC collaboration \citep{MAGIC2016NGC1275} reported the high-energy gamma-ray emission from the central source NGC1275 of the Perseus clusters, and no diffuse emission was detected.
Similarly, the upper limits provided by the LHAASO collaboration \citep{LHAASO2025ClusterUL}  also include the contribution from the central source NGC1275.

However, our results are comparable with the upper limits predicted by Fermi-LAT \citep{ackermann2014search} for massive clusters  such as A400, A3112, A1367, Coma, and EXO0422, in the local Universe. The observations reported by the SHALON experiment \citep{sinitsyna2014emission} are also comparable to our results at energies above TeV.

\begin{figure}[ht!]
\centering
\includegraphics[width=0.8\textwidth]{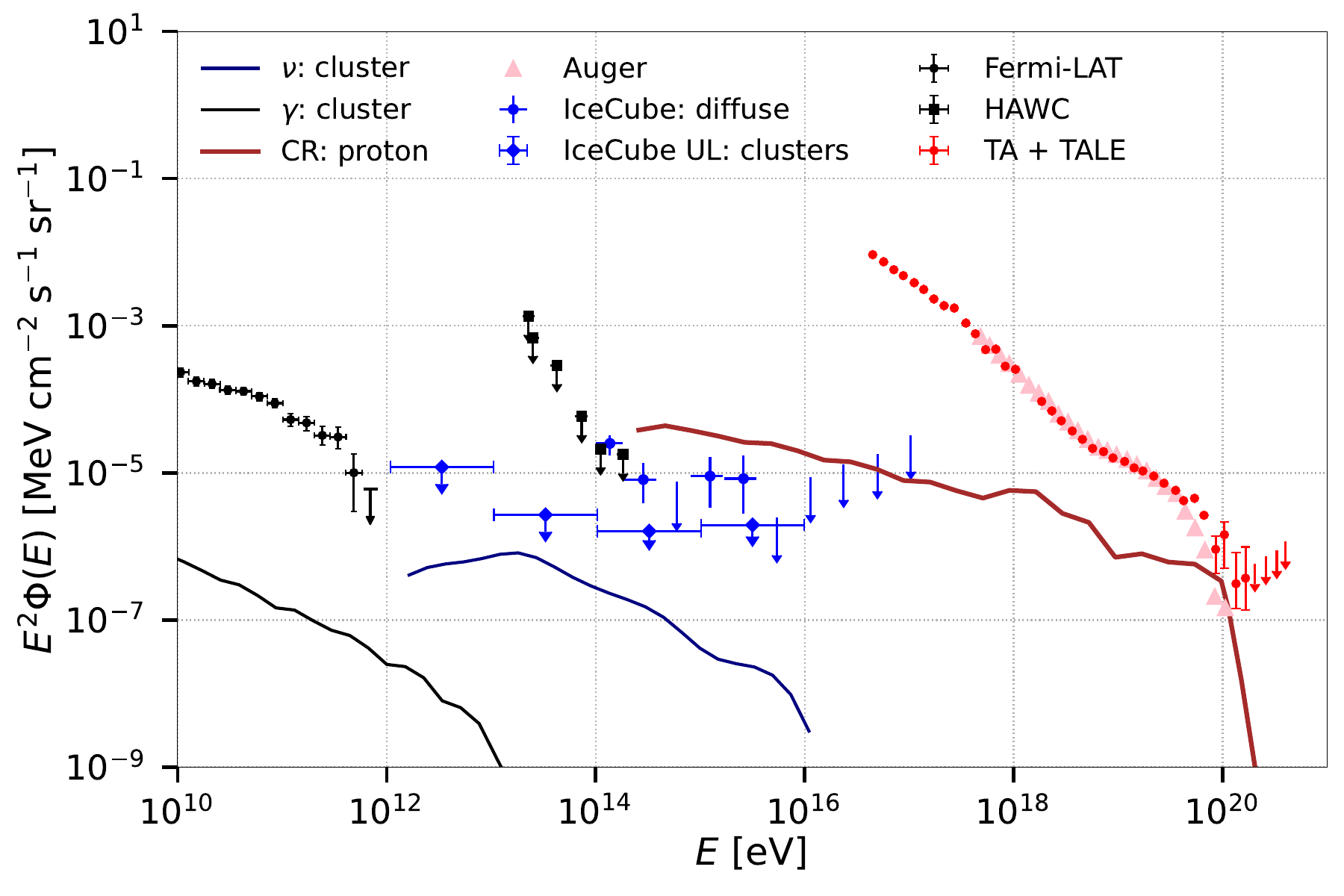}
\caption{Multi-messenger picture of Perseus cluster, we represent gamma-rays and neutrino emission. 
The black and blue lines represent the gamma-ray and neutrino spectra, respectively, for Perseus-like clusters within a distance of 80 Mpc. The results are plotted for $\alpha=2.3 \,\&\, E_\text{max}=10^{17}\, \text{eV}$.
The diffuse gamma-ray background observed by Fermi-LAT \citep{ackermann2015spectrum}, upper limits predicted by HAWC \citep{HAWC2022limits}, diffuse neutrino background observed by IceCube \citep{aartsen2015evidence, aartsen2015searches}, upper limits for clusters provided by IceCube \citep{abbasi2022searching}, and the CR spectra observed by Auger \citep{ICRC342015, halim2022constraining} and TA \citep{TACRSpectrum2015} are also depicted.}
\label{fig:Cluster_GammaNutotal}
\end{figure}

In Fig. \ref{fig:Cluster_GammaNutotal}, the total flux of gamma-ray and neutrino is shown for Perseus-like sources within a distance of $\sim 80$ Mpc.
The results are presented for clusters of mass range $\gtrsim 10^{15}\, M_{\odot}$.
The diffuse gamma-ray flux observed by Fermi-LAT \citep{ackermann2015spectrum} is higher than our results, roughly by an order of magnitude. In addition, the diffuse neutrino flux observed by IceCube \citep{aartsen2015evidence,aartsen2015searches} is quite high compared to the estimated neutrino flux in this work.
However, the upper limits provided by IceCube for galaxy clusters are comparable to our results, predicting future observations of neutrinos from these structures.  

Recently, the Telescope Array (TA) collaboration predicted an ultrahigh energy (UHE) CR source in the direction of the Perseus cluster \citep[][see also references therein]{TAPPSCUpdate, TAPPSCMagnetic2023}, and our estimations also predict the significant contribution of Perseus-like sources to CR spectra, provided that their composition is only protons. However, the Auger CR spectra  \citep{halim2022constraining} are significantly higher than our results.
We acknowledge that the intergalactic magnetic field is not considered during the propagation of CRs, which may alter our predicted CR flux.


\section{Discussion}
In this work, we investigated the high-energy gamma-ray and neutrino emissions from galaxy clusters, with a particular focus on the Perseus cluster.
The main conclusion of this work is that
galaxy clusters can contribute a sizable fraction to diffuse neutrino and gamma-ray backgrounds.
Our results for gamma-rays match well with predictions in \citep{fang2016high, fang2018linking, hussain2023diffuse, hussain2024Apj} and the neutrino flux roughly matches the existing IceCube upper limits for clusters \citep{abbasi2022searching}.
With future observations by observatories such as CTA \citep{cta2018science} and LHAASO \citep{di2016lhaaso}, we can provide more stringent constraints on gamma-ray emissions.

There have been no significant gamma-ray and neutrino observations from clusters, yet.
MAGIC collaboration has observed the gamma-ray emission from NGC 1275 above $100$ GeV of energy but did not detect any diffuse gamma-rays from the Perseus cluster \citep{MAGIC2016NGC1275}.
More recently, LHAASO has provided upper limits for the Perseus cluster, but these estimates also include the contribution from the central source NGC1275.
Our results for gamma-rays are quite below the LHAASO and MAGIC predictions, maybe because these estimates are more consistent with the point source emission from NGC1275, but we follow the gas density profile of clusters to inject CRs and their subsequent production of gamma-rays and neutrinos.

Following the sensitivities of CTA and LHAASO, our results predicted the future observation of gamma-rays from the Perseus cluster by experiments such as CTA and LHAASO.
In addition, the neutrino flux we computed for Perseus-like clusters is quite comparable to the existing IceCube upper limits, predicting the future observations of neutrinos from galaxy clusters by observatories such as IceCube, the Cubic Kilometre Neutrino Telescope (KM3NeT), and the Giant Radio Array for Neutrino Detection (GRAND). 
Furthermore, our results for CRs are comparable with the TA data, but well below the CR flux reported by the Auger observatory \citep{halim2022constraining}.

Our estimates might be subject to change if we modify parameters, such as considering the mix CR composition, including heavy ions instead of only protons. Also, the intergalactic magnetic field may have a slight impact on our results. 
On the other hand, given the large uncertainty in the intergalactic magnetic field, combining our detailed modeling with multi-messenger observations will enable us to constrain its properties, especially by establishing an upper limit.
Developing more detailed modeling, including the effects of mixed CR composition and intergalactic magnetic field, is one of our future endeavors.

However, parameters chosen in our study are well established and similar to previous studies \citep{fang2018linking, hussain2021high, hussain2023diffuse}. Furthermore, our studies establish a clear connection among gamma-rays, neutrinos, and CRs, enabling us to investigate the interplay between CRs and gas in ICM.
This work is crucial for understanding the microphysics of ICM.

\paragraph{Acknowledgment}
SH acknowledges support from the COFUND action of Horizon Europe’s Marie Sklodowska-Curie Actions research programme, Grant Agreement 101081355 (SMASH).
The simulations in this work were carried out using the VEGA high-performance computing (HPC) system.
We are grateful for the support of the Slovenian National HPC initiative and EuroHPC JU under the project "VEGA - National Supercomputing Centre".




\end{document}